\documentclass[prd,twocolumn,showpacs,preprintnumbers]{revtex4}
\pdfoutput=1
\usepackage{amsfonts,amsmath,amssymb}
\usepackage{mhchem}
\usepackage{graphicx}
\usepackage{color}
\usepackage{natbib}
\usepackage{lipsum}
\usepackage[plainpages=false, colorlinks=true, anchorcolor=blue, linkcolor=blue, citecolor=blue, bookmark
s=false]{hyperref}

\newcommand{\rthis}[1]{\textcolor{black}{#1}}

\begin{document}

\newcommand{\apjl}{Astrophys. J. Lett.}
\newcommand{\apjs}{Astrophys. J. Suppl. Ser.}
\newcommand{\aap}{Astron. \& Astrophys.}
\newcommand{\aj}{Astron. J.}
\newcommand{\araa}{Ann. Rev. Astron. Astrophys. } 
\newcommand{\mnras}{Mon. Not. R. Astron. Soc.}
\newcommand{\solphys}{Solar Phys.}
\newcommand{\jcap}{JCAP}
\newcommand{\pasj}{PASJ}
\newcommand{\pasa}{Pub. Astro. Soc. Aust.}
\newcommand{\apss}{Astrophysics \& Space Science}
\newcommand{\aaps}{Astron. Astrophys. Suppl. Ser.}
\newcommand{\Dtwoo}{$\mathrm{D_2O}$ }
\newcommand{\cl}{$\mathrm{^{36}Cl}$ }

\title{Generalized Lomb-Scargle analysis of   $\rm{^{36}Cl}$ decay rate measurements at PTB and BNL}

\author{Akanksha Dhaygude}
\altaffiliation{ep15btech11005@iith.ac.in}
\author{Shantanu  Desai}
\altaffiliation{shntn05@gmail.com}
\affiliation{Department of Physics, IIT Hyderabad, Kandi, Telangana 502285 India}


%

\begin{abstract}
Recently Pomme et al.~\cite{Pommesolar} did an analysis of $\ce{^{36}Cl} $ radioactive decay data from 
measurements at the Physikalisch-Technische Bundesanstalt  (PTB), in order to verify the claims by  Sturrock and collaborators of an influence on beta-decay rates measured at Brookhaven National Lab (BNL) due to the rotation-induced modulation of the  solar neutrino flux.  Their analysis excluded any sinusoidal modulations in the frequency range from 0.2-20/year.
We carry out an independent analysis of the same PTB and BNL data,   using the generalized Lomb-Scargle periodogram to look for any statistically significant peaks in the range from 0 to 14  per year, and by evaluating the significance of every peak using multiple methods. Our results for the PTB data  are in agreement with those by Pomme et al.  For BNL data, we do find peaks at  some of the same frequencies as Sturrock et al., but the significance is much lower. All our analysis codes and datasets have been made publicly available. 
\end{abstract}
                                                     
\maketitle

\section{Introduction}

Sturrock and collaborators have argued in a number of works  over more than a decade (eg. Refs.~\cite{Sturrock13,Sturrock16,Sturrock18}  and references therein) that beta decay rates for  a large number of nuclei  exhibit variability  and show periodicities at multiple frequencies, some of which  have been associated with solar rotation as well as other processes in the solar core.  They have also found similar peaks at 12.7 per year in the Super-K solar neutrino flux (from the first five years of data)~\cite{Sturrock01}, which they have argued to be due effects  of solar rotation.  Furthermore,  they have correlated  the two sets of findings,  and argued for an influence of solar neutrinos on beta-decay rates.

However, many other groups have failed to reproduce the periodicities in the beta-decay  results, while analyzing the same data as well as decays of the same elements from other experiments. A review of some of these claims and rejoinders can be found in Refs.~\cite{Sturrock16,Kossert,Pomme}. In our previous works, we have also carried out an independent analysis of some of these claims and  found evidence of periodicities at some of the same frequencies as found by Sturrock et al., albeit with a lower significance~\cite{Desai16,Tejas}.

In this work, we focus on addressing  the claimed periodicities  in the beta decay rates of $^{36}$Cl.  Sturrock et al.~\cite{Sturrockcomp} have argued for periodicities with periods at 1/year  and 12.7/year  (or 28.7 days)  in the $^{36}$Cl  decay rates of the Brookhaven National Lab~\cite{BNL}  counting experiment. They have argued that the peak at 12.7/year  is  indicative of the synodic rotation rate of the radiative zone of the Sun, since it
matches the value of   28.7 days determined using helioseismology~\cite{Schou,Komm}.  These results were rebutted by Pomme et al.~\cite{Pommesolar}, who found  no evidence for periodicities in the decay rates of $^{36}$Cl using more accurate measurements at Physikalisch-Technische Bundesanstalt Braunschweig (PTB), obtained using the triple-to-double coincidence ratio measurement techniques~\cite{Kossert}. Pomme~\cite{Pomme15,Pomme16} has also raised concerns about the detector stability and control of experimental uncertainties in the BNL measurements, which are now more than three decades old.  Furthermore, the invariability of the decay constants for \cl was also demonstrated using triple-to-double coincidence ratio measurements~\cite{Kossert14}, to refute claims of oscillations ascribed to the changes in Earth-Sun distance.

In this work, we independently try to adjudicate the conflicting results between these two works by doing an independent analysis of the beta decay residual data from both the  BNL and PTB measurements (which were kindly provided to us by  S.~Pomme) using the Generalized Lomb-Scargle periodogram~\cite{Lomb,Scargle,Kurster}. We search the frequency range between 0-14 /year (or up to 26 days), since this covers the  frequency range  associated with  solar rotation~\cite{Sturrockcomp,Sturrocksolar}. We calculate the significance of the peaks using all the available methods provided in the {\tt astropy}~\cite{astropy} library used to calculate the Lomb-Scargle periodogram.

The outline of this paper is as follows. We briefly recap some details of the Lomb-Scargle periodogram and different methods of calculating the $p$-value in Sect.~\ref{sec:ls}. A summary of the results by Sturrock and collaborators and the re-analysis by Pomme and collaborators is discussed in Sect.~\ref{sec:prevanalysis}. 
Our analysis of the PTB and BNL datasets  is described in Sect.~\ref{sec:analysis}. 
We conclude in Sect.~\ref{sec:conclusions}.

\section{Generalized Lomb-Scargle Periodogram}
\label{sec:ls}
The Lomb-Scargle (L-S)~\cite{Lomb,Scargle} (see Ref.~\cite{Vanderplas, astroml} for recent extensive reviews) periodogram is a widely used technique to look for  periodicities in unevenly sampled datasets. The main goal of the L-S periodogram  is to determine the  frequency ($f$)  of a  periodic signal in a time-series dataset  $y(t)$
given by:
\begin{equation}
y(t)=a\cos(2\pi f t)+ b \sin(2 \pi f t).
\label{eq:yt}
\end{equation}
The L-S periodogram calculates the power as a function of frequency, from which one needs to infer the presence of a sinusoidal signal and assess the significance.

For this analysis, we use a modified version of the  L-S periodogram  proposed by Zechmeister and Kurster~\cite{Kurster}, which is known in the literature as the generalized L-S periodogram~\cite{Bretthorst,Kurster} or the floating mean periodogram~\cite{Cumming,Vanderplas15,Vanderplas} or the Date-Compensated Discrete Fourier Transform~\cite{Ferrazmello}. The main change in this method is that an arbitrary offset is added to the mean values.   More details about this method and comparison with the normal L-S periodogram  are discussed in Refs.~\cite{Bretthorst,Kurster,Vanderplas,astroml,proc} and references therein.

For any  sinusoidal modulations at a given frequency, one would expect a peak in the L-S periodogram. To assess the  statistical significance of such a  peak, we need to calculate its  false alarm probability (FAP) or $p$-value. A plethora of methods
have been developed  to estimate the FAP of peaks in L-S periodogram,  ranging from analytical methods~\cite{Scargle} to Monte-Carlo simulations~\cite{Suveges}. We enumerate the different methods used to calculate the FAP for our analysis. All of  these can be implemented using the {\tt astropy} package, which we used in this work.
\begin{itemize}
    \item {\bf Baluev}

    This method implements the approximation proposed by Baluev~\cite{Baluev}, which uses extreme value statistics for stochastic process, to compute an upper-bound of the FAP for the alias-free case. Their analytical formula for the FAP  can be found in Refs.~\cite{Baluev,Vanderplas}.
    
    \item {\bf Bootstrap}

    This method uses non-parametric bootstrap resampling  as described in Ref.~\cite{Vanderplas}. Effectively, it computes many L-S periodograms on simulated data at the same observation times. The bootstrap approach can very accurately determine the false alarm probability, but is very computationally expensive.
    
    \item {\bf Davies}

    This method is related to the Baluev method, but loses accuracy at large false alarm probabilities, and is described in Ref.~\cite{Davies}.
    \item {\bf Naive}
    
        This method is a simplistic method, based on the assumption that well-separated areas in the periodogram are independent.
    The total number of such independent frequencies depend on the sampling rate and total duration  and is explained in Ref.~\cite{Vanderplas}. 
    \end{itemize}

Once the FAP is calculated, one can convert this FAP to a $Z$-score or  significance in terms of number of sigmas. This is traditionally estimated from the number of
standard deviations that a Gaussian variable would fluctuate in one direction to give the corresponding FAP~\cite{Cowan11,Ganguly}.

\section{Recap of results by Sturrock  and Pomme  }
\label{sec:prevanalysis}
Here, we briefly summarize the analysis in Sturrock et al.~\cite{Sturrockcomp} (S16 hereafter) and Pomme et al.~\citep{Pomme} (P17, hereafter). S16 analyzed the $\ce{^{36}Cl}$ decay data from BNL.  The data was detrended   and normalized to account for the exponential decay. Power spectrum of this detrended data, based on the procedure outlined in Ref.~\cite{Sturrock03}  was used to search for periodicities at different frequencies.  The maximum power was found at a frequency of 1 /year, corresponding to a $p$-value of $2.7 \times 10^{-7}$.  This $p$-value was computed using the Press-Bahcall shuffle test~\cite{Bahcall91}.

Since the $\ce{^{36}Cl}$ decay data was found to be non-uniform, the same data was analyzed using  spectrograms and phasegrams,  in order to look for transient oscillatory cycles. From this, it was discerned that for $\ce{^{36}Cl}$, annual modulation was conspicuous between 1984 and 1986, but later switched off.
They further analyzed the data to look for evidence of solar rotation. The range of frequencies, which they scanned based on the observed synodic rotation~\citep{Sturrock06} corresponds to 9-14 per year or periods between 26-41 days. 
From these spectrograms, S16 found  evidence for oscillations with a frequency of 12.7/ year, which is compatible with a source in the solar radiative zone. The $p$-value  of this peak was also obtained using the shuffle test, and found to be $1.5 \times 10^{-5}$ in a band of unit width. Accounting for the look-elsewhere effect by incorporating the bandwidth of 5 per year, the $p$-value was found  to be $7.5 \times 10^{-5}$. S16 then discuss the correlation between these peaks and peaks near this same frequency found in Super-K solar neutrino data~\cite{Sturrock03}.

P17 applied the generalized (or floating-mean) Lomb-Scargle periodogram\cite{Kurster,Vanderplas} to both the BNL and PTB datasets in order to look for periodicties between 0 and 20 per year. They also looked for concurrent peaks in PTB, BNL, and Super-K solar neutrino datasets. They do not  find any peak at 9.43/year (seen in the Super-K data) in the BNL or PTB data. Although they found many  significant peaks in the BNL data at 11 and 12.7/year, none of these peaks were visible at the same frequencies in the PTB data. P17 then fit the BNL and PTB  data  to sinusoids at frequencies of 9.43, 11.0, and 12.7 /year. The amplitudes in the PTB data found are $\mathcal{O}(10^{-3})$\%, statistically indistinguishable from zero and about an order of magnitude lower than the amplitudes reported in S16. Therefore, they disagree with the conclusions in S16.

\section{Analysis and results}
\label{sec:analysis}
We first recap the input datasets used for our analysis, then  describe the L-S analysis procedure, and finally present our results.
\subsection{BNL and PTB Datasets}

\label{dataset}
The BNL dataset comprises of 364 measurements formed from countings of $\ce{^{36}Cl} $ and $\ce{^{32}Si}$ decays in gas flow proportional counter at BNL~\cite{BNL}, over the period of 1982 to 1990.  More details of these measurements can be found in Ref.~\cite{BNL}.    For  each  nuclei,  the daily decay rate was obtained by averaging over 20 measurements.  The data was detrended and normalized to account  for  the  exponential  decay. Experimental uncertainties used in these measurements are discussed in Refs.~\cite{Pomme15,Pomme16}.
The exact total duration of this dataset  is 7.83 years and the median sampling interval is 0.00279 years or approximately 1 day. 
We use an uncertainty of  0.13\%  for every data point. \rthis{These uncertainties are same as those used in P17 and S16}. The raw BNL decay data (after removing the exponential dependence) as a function of time can be found in Fig.~\ref{fig3}.

\begin{figure*}
\includegraphics[scale=0.5]{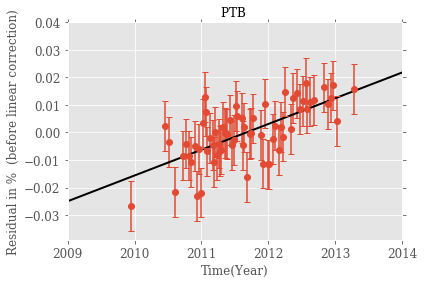}
\caption{The original PTB dataset~\cite{Pomme} showing the $\ce{^{36}Cl}$ decays. The best-fit line has  a slope equal to $9.36 \times 10^{-5}$ and $y$-intercept is equal to 0.81.}
\label{Figi}
\end{figure*}

\begin{figure*}
\includegraphics[scale=0.5]{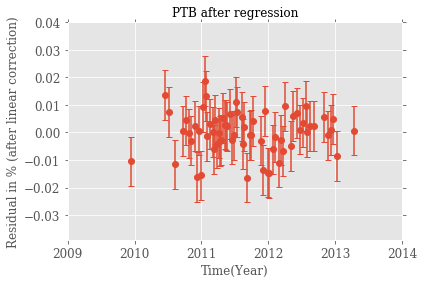}
\caption{Residuals in the PTB dataset after applying the  linear correction (outlined in Fig.~\ref{Figi}).  The uncertainty for each data point is  $0.009$\%.}
\label{Fig2}
\end{figure*}

\begin{figure*}
\includegraphics[scale=0.5]{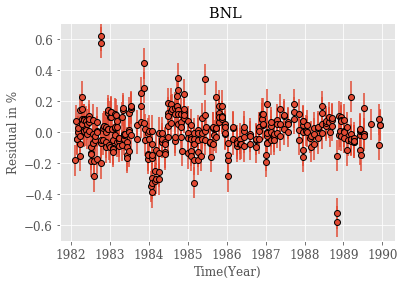}
\caption{Relative difference from the mean value of measured $\ce{^{36}Cl}$  at the BNL (applying $0.13\%$ uncertainty on individual data)~\cite{BNL}.}
\label{fig3}
\end{figure*}

The PTB experiment consists of liquid scintillation vials with $\ce{^{36}Cl}$ in solution, which were prepared in December 2009.  The decays were  measured 66 times between December 2009 and  April 2013 in the custom-built TDCR detector at the PTB. More details of the PTB experiment and the setup used for these measurements can be found in Ref.~\cite{Kossert14}. 
The exact total duration of the PTB dataset is 66 days and the median sampling time is equal to 0.0328 years or approximately 12 days. 
The raw PTB data is shown in Fig.~\ref{Figi}. 
The uncertainty in each data point was 0.009\%. The same uncertainity was used in P16.
Linear correction was applied to the dataset to compensate for a long-term instability due to increasing colour quenching in the scintillation cocktail with time~\cite{Pommesolar,Kossert10}. For this, {\tt numpy.polyfit} function is used to apply a  linear correction~(Fig.~\ref{Figi}), and the residuals after applying this linear correction are shown in Fig.~\ref{Fig2}.

\begin{figure}
\includegraphics[scale=0.6]{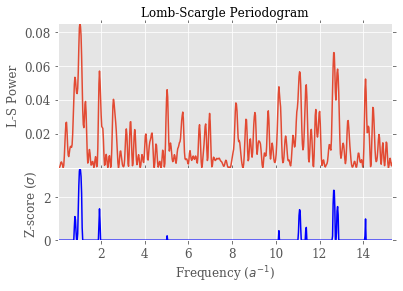}
\caption{Weighted L–S periodograms of $ \ce{^{36}Cl} $ decay rate data measured at the BNL (top panel) for  frequencies in the range 0 - 14 $ a^{-1}$). The bottom panel shows the Z-scores greater than zero  (obtained from $p$-value calculated via the Baluev method) for different frequencies.}
\label{fig5}
\end{figure}

\begin{figure}
\includegraphics[scale=0.6]{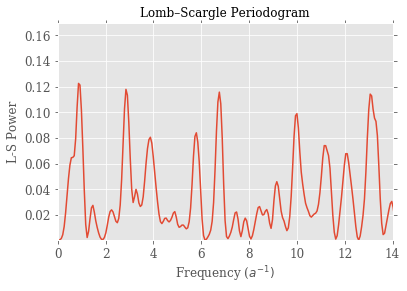}
\caption{Weighted L–S periodograms of $ \ce{^{36}Cl} $ decay rate data measured at the PTB for frequencies in the range 0 - 14 $ a^{-1}$). The Z-scores are negligible and not shown here.}
\label{fig6}
\end{figure}
%



\subsection{Power spectrum analysis}
\label{sec:FAP}
For our analysis, we applied the generalized L-S periodogram as described in Sect.~\ref{sec:ls}. We used the {\tt astropy}~\cite{astropy} implementation of the  L-S periodogram.
 These periodograms can be found  Figs.~\ref{Fig2}  and Fig.~\ref{fig3}  for BNL and PTB data,  respectively.  We normalized the periodogram by the residuals of the data around the constant reference model. With this normalization, the power varies between 0 and 1. The same normalization has been used in P17. However, S16  (and also  one of our earlier works~\cite{Desai16}) used the normalization proposed by Scargle~\cite{Scargle}. The relation between these two normalizations  can be found in Ref.~\cite{Tejas}. 
 
 For our analysis,  we report results for   frequencies from 0 to 14 per year. This frequency range covers    the sweet spot for  solar rotation-related phenomenon (from 8-14  per year) and is also sensitive to lower frequencies (such as 1/year), which were found to be significant in S16. The corresponding  Nyquist frequency for the two data sets is equal to  132/year for BNL and  15/year for PTB. Therefore, the  \rthis{frequency range}  which we have searched for is \rthis{much smaller}  than the Nyquist frequency. We also searched for peaks at higher frequencies (up to the Nyquist limit). \rthis{But  none of them were found to be significant in agreement with  previous analysis by Sturrock and collaborators). }
 The frequency resolution used in the periodograms was 0.025 per year  and 0.06 per year for BNL and PTB respectively.  This resolution is determined from the reciprocal of the total duration of the dataset.  Our  results on the power spectrum  for the generalized L-S periodiogram for  both BNL  and PTB can be found in Fig.~\ref{fig5} and Fig.~\ref{fig6} respectively. 

While considering these frequencies and corresponding power, we need to identify the significance of each peak.  This significance is usually determined from the FAP
and Z-score. We calculated the FAP (along with associated Z-score) using all the  four methods discussed in Sect.~\ref{sec:ls}.  A tabular summary of the powers at some of the largest  peaks along with FAP and  Z-score at each of these frequencies can be found in Tables~\ref{BNL} and ~\ref{PTB} for BNL and PTB respectively.   \rthis{For BNL data, we have also shown the Z-score obtained using the Baluev FAP in Fig.~\ref{fig5}.}

\subsection{Results}
We now discuss below the results for each of the two datasets:
\begin{itemize}
\item {\bf BNL}. The BNL L-S power spectrum is shown in Fig.~\ref{fig5}. We find peaks with power $> 0.05$ at frequencies (less than 14 /year) corresponding to  1.04, 1.93, 1.08, 11.08 12.65, and 12.82  per year. The powers and FAP of these peaks are tabulated in Table~\ref{BNL}. These peaks have also been found to be significant in S16. However, the FAPs which we get are about two  to three orders of magnitude smaller than in S16. The FAPs calculated for each frequency using the multiple methods are of the same order of magnitude. All Z-scores, which we obtained are less than 5$\sigma$, (a traditional threshold used for deeming something as a new discovery~\cite{Lyons}).
The smallest FAP which we get is close to 1 per year with a FAP of  $\mathcal{O}$ ($10^{-4}$), corresponding  to Z-score between 3.3$\sigma$ and 3.8$\sigma$. This frequency also had the maximum significance in S16.  For frequencies in the range from 8 to 14/year, the minimum FAP is at a frequency of  12.65/year, corresponding to a FAP of  $\mathcal{O}$ ($10^{-3}$), with a significance of 2.8$\sigma$. This is the closest frequency to  12.7/year, found to be interesting  in S16. Therefore, although we do see peaks at some of the same frequencies as seen in S16, the
FAPs, which we obtain are about two to three orders of magnitude larger than in S16.
Therefore, the significance we obtained for these peaks is marginal and not as large as in  S16.
\item {\bf PTB} The PTB L-S power as a function of frequency can be found in Fig.~\ref{fig6}, and a tabular summary of the powers and FAP of the most significant peaks  (with power $>0.1$) can be found in Table~\ref{PTB}. We see that none of the FAPs have values less than 0.1, implying that all of them are consistent with  statistical fluctuation, and there is no periodicity at any of the frequencies  found to be significant in S16. Therefore, we concur with the findings of P17. If the peaks found in S16 were indicative of a solar influence, similar periodicities should have been seen in the PTB data.  We do not find any such evidence. 
\end{itemize}

\vspace{3mm}
\begin{table*}
\begin{tabular}{|c|c|c|c|c|c|}

 \hline
 Frequency($a^{-1}$) & L-S Power & FAP : Baluev  & FAP : Davies  &FAP : Naive    & FAP : Bootstrap  \\
 \hline
 1.04&	0.09&	$3.6 \times 10^{-4}$ (3.38$\sigma$)&$8.5 \times 10^{-5}$ (3.76$\sigma$)&0.001 (3.09$\sigma$)&	$3.6 \times 10^{-4}$ (3.38$\sigma$)\\
 
 1.93&	0.06&	0.07 (1.48$\sigma$)&0.02 (2.64$\sigma$)	&0.05 (1.64$\sigma$)&	0.07 (1.48$\sigma$)	\\
 
 11.08&	0.06&	0.08 (1.41$\sigma$)&	0.02 (2.64$\sigma$)&	0.05 (1.64$\sigma$)&	0.08 (1.41$\sigma$)\\
 
 12.65&	0.07&	0.01 (2.33$\sigma$)&	$2.68 \times 10^{-3}$ (2.78$\sigma$)&	0.01 (2.33$\sigma$)&	0.01(2.33$\sigma$)\\
 
 12.82&	0.06&	0.06 (1.55$\sigma$&	0.02 (2.64$\sigma$)&	0.04 (1.73$\sigma$))&	0.06 (1.55$\sigma$)\\
 
14.10&0.05 &	0.16 (0.99$\sigma$)  &	0.05 (1.64$\sigma$)&	0.12 (1.17$\sigma$)&	0.17 (0.95$\sigma$)\\
 \hline
\end{tabular}
\caption{L-S powers and FAP for BNL data using multiple methods:  Baluev,  Davies, Naive, Bootstrap methods, for only those frequencies for which the  power is greater than 0.05.  The numbers  in parenthesis represent the Z-score, found using the method prescribed in Ref.~\cite{Cowan11}. We find peaks at some of the same frequencies found to be statistically significant in S16, but our FAPs are about 2-3 orders of magnitude higher than those found in S16. }
\label{BNL}
\end{table*}

\begin{table*}
\begin{tabular}{|c|c|c|c|c|c|}

 \hline

 Frequency($a^{-1}$)& L-S Power & FAP : baluev& FAP : davies&FAP : naive    & FAP : bootstrap  \\
 \hline
 0.87&0.12	&0.99&	4.39&	0.93&	0.99\\
2.85&0.12	&	0.99&	5.07&	0.96&	0.99\\
 6.76& 0.12	&0.99&	5.42&	0.97&	1\\
13.07& 0.11 &	0.99&	5.68&	0.97&	1\\

 \hline
\end{tabular}
\caption{ PTB L-S powers along with the FAP for the top four most significant peaks  (with power $>0.1$)  computed using the same methods as in Table~\ref{BNL}. We find that none of the FAPs are less than 0.1, and hence all the peaks discussed here are consistent with noise. \rthis{Therefore, no Z-scores are reported in this table}.}
\label{PTB}
\end{table*}





 


\section{Conclusions}
\label{sec:conclusions}
The aim of this work was  to adjudicate  the controversy between two groups (S16 and P17) regarding the periodicities  in nuclear beta decay rates of $\ce{^{36}Cl}$, and possible solar influence on these potential periodic decay rates. 

For this purpose, we independently re-analyzed both the BNL data, for which S16 found evidence for statistically significant peaks at multiple frequencies as well as the PTB data, for which P17 could not find any corroborative evidence  at the same frequencies as in S16. We have used the  generalized or floating-mean  L-S  periodogram~\cite{Kurster} (similar to our previous works, where we analyzed the Super-K solar neutrino~\cite{Desai16} and $\ce{^{90} Sr/^{90}Y}$  decay data~\cite{Tejas}), to look for periodicities in the frequency range from 0 to 14  per year, since this is the same frequency range, wherein which S16 found periodicities. 

When we analyzed the BNL data, we found peaks in the L-S periodogram at mostly the same frequencies found to be significant in S16. However, the FAP which we found is about 2-3 orders of magnitude larger than S16. Therefore, according to our analysis, none of the peaks found in S16 are statistically significant \rthis{(for discovery at 5$\sigma$ significance)} and indicative of any solar or any other external influence.  \rthis{The maximum significance we find in the BNL data is for a frequency of 1 year, corresponding to a significance of   3.8$\sigma$. The significance for the frequency close to 12.7 /year  (within the range of solar rotation) is about 2.8$\sigma$. Although, the cause of  this peak is not investigated in this work, there is no evidence that this related to solar rotation. This could have a  more prosaic explanation, related to  systematics of the BNL detector setup or failure to control the experimental uncertainties~\cite{Metro,Barrow,Pommesolar,Pomme15,Pomme16}}.

However, when we analyzed the more recent and high-precision PTB data \rthis{(whose stability is about an order of magnitude more stringent than the BNL measurements)}, we do not find any peaks with FAP$<0.1$, indicating that the $\ce{^{36}Cl}$
PTB decay data contain no periodicities. Therefore, we agree with the conclusions in P17 regarding this dataset.

To promote transparency in data analysis, we have made our analysis codes and data available online~\cite{Dhaygude}. These can be easily applied to look for  periodicities in other datasets.

\begin{acknowledgements}
We are grateful to Stefaan Pomme for providing us the data for the PTB  and BNL measurements analyzed  in P17 and useful correspondence.
\end{acknowledgements}

\bibliography{lombscargle}

\end{document}